\UseRawInputEncoding
\documentclass[prl,twocolumn,preprintnumbers,amsmath,amssymb]{revtex4}

\usepackage{graphicx}
\usepackage{dcolumn}
\usepackage{bm}
\usepackage{color}
\usepackage{epstopdf}

\def\eps{\varepsilon}

\newcommand{\br}{ {\bm r}}

\def\bsigma{{\boldsymbol{\sigma}}}

\begin{document}



\title{Energy and wave-action flows underlying Rayleigh-Jeans thermalization\\
 of optical waves propagating in a multimode fiber}

\author{K. Baudin$^{1}$, A. Fusaro$^{1}$, J. Garnier$^{2}$, N. Berti$^{1}$, K. Krupa$^{3}$, I. Carusotto$^{4}$, S. Rica$^{5}$, G. Millot$^{1,6}$, A. Picozzi$^{1}$}
\affiliation{$^{1}$ Laboratoire Interdisciplinaire Carnot de Bourgogne, CNRS, Universit\'e Bourgogne Franche-Comt\'e, Dijon, France}
\affiliation{$^{2}$ CMAP, CNRS, Ecole Polytechnique, Institut Polytechnique de Paris, 91128 Palaiseau Cedex, France}
\affiliation{$^{3}$ Institute of Physical Chemistry Polish Academy of Sciences, 01-224 Warsaw, Poland}
\affiliation{$^{4}$ INO-CNR BEC Center and Dipartimento di Fisica, Universit\`a di Trento, I-38123 Povo, Italy}
\affiliation{$^{5}$ Universidad Adolfo Ib\'a\~nez, Pe\~nalol\'en, 7910000 Santiago, Chile}
\affiliation{$^{6}$ Institut Universitaire de France (IUF), 1 rue Descartes, Paris, France}


\begin{abstract}
The wave turbulence theory predicts that a conservative system of nonlinear waves can exhibit a process of condensation, which originates in the singularity of the Rayleigh-Jeans equilibrium distribution of classical waves.
Considering light propagation in a multimode fiber, we show that light condensation is driven by an energy flow toward the higher-order modes, and a bi-directional redistribution of the wave-action (or power) to the fundamental mode and to higher-order modes.
The analysis of the near-field intensity distribution provides experimental evidence of this mechanism. 
The kinetic equation also shows that the wave-action and energy flows can be inverted through a thermalization toward a negative temperature equilibrium state, in which the high-order modes are more populated than low-order modes.
In addition, a Bogoliubov stability analysis reveals that the condensate state is stable.
\end{abstract}

\pacs{42.65.Sf, 05.45.a}

\maketitle

Bose-Einstein condensation (BEC) has been predicted and experimentally reported in genuine quantum systems, such as quantum degenerate gases of ultracold atoms \cite{stringari}, exciton polaritons \cite{carusotto13}, magnons \cite{demokitrov06} and photons \cite{weitz,fischer19}. 
On the other hand, several studies based on the wave turbulence theory \cite{zakharov92,newell01,nazarenko11,Newell_Rumpf,shrira_nazarenko13} predict that nonlinear waves can also exhibit a phenomenon of condensation \cite{newell01,nazarenko11,nazarenko05,PRL05,onorato06,berloff07,PD09,brachet11,PRA11b,
laurie12,Fleischer,suret,PR14,nazarenko14,rj_cond_magnons15,cherroret15,PRL18}.
Although the physics of quantum gases and wave condensation are different, the underlying mathematical origin of the condensation process is similar because of the 
common low-energy divergence of the equilibrium Bose distribution for quantum particles and the equilibrium Rayleigh-Jeans (RJ) distribution for waves \cite{PRL05,nazarenko11}. 
Other forms of condensation processes have been discussed for optical cavity systems, whose nonequilibrium forced-dissipative features\cite{carusotto13,conti08,berloff13,fischer14,fischer16,turitsyn12,turitsyn15} lead to different forms of universal properties \cite{zamora17}.

Full 3D quantum thermalization and condensation with optical waves in a conservative cavity-less free propagation geometry has been predicted in \cite{chiocchetta16}, but has also been anticipated to require prohibitive large propagation lengths. 
Reducing to an effective 2D geometry using monochromatic classical light helps observing condensation effects, but still requires propagation lengths that challenge experimental feasibility \cite{PRL18}. 
Actually, thermalization to the RJ equilibrium is not even properly defined when the optical beam propagates in a bulk medium because of the ultraviolet catastrophe inherent to classical optical waves.
This issue can be circumvented by considering a waveguide configuration, whose finite number of modes regularizes the ultra-violet catastrophe and also substantially reduces the rate of thermalization \cite{PRA11b,PR14}.
In this respect, a remarkable phenomenon of spatial beam self-organization, termed `beam self-cleaning', has been recently discovered in (graded index) multimode fibers (MMF) \cite{krupa16,wright16,liu16,krupa17}.
Recent works suggested that this phenomenon of beam self-cleaning can be interpreted as a consequence of a wave thermalization and condensation process \cite{PRL19,pod19,PRA19,christodoulides19,kottos20}.
In particular, a wave turbulence kinetic equation (KE) describing this effect has been derived in \cite{PRL19,PRA19}.
This process has been experimentally demonstrated in a recent work \cite{baudin_arxiv}, 
where the condensate fraction across the transition to condensation has been found in agreement with the RJ equilibrium theory.

Our aim in this article is to provide more physical insights into the experimental results reported in \cite{baudin_arxiv}.
We recall in this respect that wave condensation is usually understood as an inverse turbulence cascade that increases the level of nonlinearity at large scales (i.e. low wave-numbers), up to a breaking point of the weak turbulence theory \cite{nazarenko11}. 
In the {\it focusing} regime of our experiment, such a nonlinear breaking point is usually regularized by the (Benjamin-Feir) modulational instability, which leads to the generation of coherent soliton-like structures (`soliton condensation') \cite{nazarenko11,laurie12,rumpf01,zakharov04,rumpf_zakh09}. 
At variance with this {\it strongly nonlinear process that occurs far from thermal equilibrium}, in our experiments the transition to condensation is driven by the thermalization to the RJ equilibrium in the {\it weakly nonlinear regime}. 
More precisely, we show that the process of condensation is characterized by a flow of the energy toward the higher-order modes, and a bi-directional redistribution of the wave-action (or optical power, or particle number in a corpuscular picture), from intermediate modes to both the fundamental and the higher-order modes.


{\it Modal nonlinear Schr\"odinger equation.--}
We consider the (2D+1) nonlinear Schr\"odinger equation (NLSE) accounting for the polarization degree of freedom, which is known to describe the transverse spatial evolution of an optical beam in a waveguide modelled by a confining potential $V(\bm r)$ [with $\bm r=(x,y)$] \cite{PR14}.
Following the experiments of beam cleaning, we consider a parabolic shaped potential $V(\bm r)$ modelling a graded-index MMF, with the mode eigenvalues $\beta_p=\beta_0(p_x+p_y+1)$ (the index $p$ labels the two integers $(p_x, p_y)$ that specify a mode), where $\beta_0=1/(n_c k_0 r_o^2)$ with $k_0=2\pi/\lambda$, $r_o$ the radius of the fundamental mode, $\lambda$ the laser wavelength, and $n_c$ the core refractive index.
By expanding the random wave into the normalized Hermite-Gauss modes ($u_p(\br)$) of the MMF, the modal NLSE for the evolutions of the vector modal components ${\bm a}_p=(a_{p,x},a_{p,y})^T$ reads \cite{PRL19,PRA19}:
\begin{eqnarray}
i \partial_z {\bm a}_p = \beta_p {\bm a}_p + {\bf D}_p(z) {\bm a}_p - \gamma  {\bm P}_p({\bm a}),
\label{eq:nls}
\end{eqnarray}
where the nonlinear terms read
${\bm P}_p ({\bm a})  = \sum_{l,m,n} S_{plmn} \Big(\frac{1}{3}
{\bm a}_l^T {\bm a}_m {\bm a}_n^* +\frac{2}{3} {\bm a}_n^\dag  {\bm a}_m {\bm a}_l\Big)$,
$S_{plmn}$ 
denoting the overlap among the modes  -- note that $S_{0000}=1$ \cite{PRA19}.
To explain the experiments of beam-cleaning, it is important to introduce the impact of a structural disorder, which is known to affect light propagation in MMFs due to inherent imperfections and external perturbations \cite{kaminow13}.
We consider in (\ref{eq:nls}) the dominant contribution of weak disorder.
In its most general form that conserves the wave-action $N=\sum_p |{\bm a}_p|^2$, the Hermitian matrices ${\bf D}_p(z)$ are expanded into the Pauli matrices $\bsigma_j$, ${\bf D}_p(z) = \sum_{j=0}^3 \nu_{p,j} (z) \bsigma_j$, where $\bsigma_0$ is the identity matrix and $\nu_{p,j}(z)$ are independent and identically distributed real-valued random processes, with variance $\sigma^2$ and correlation length $\ell_c$. 
Introducing the parameter $\Delta \beta = \sigma^2 \ell_c$, the characteristic length scale of disorder is $L_{d}=1/\Delta \beta$ \cite{PRA19}.
Finally note that since the disorder is (``time") $z$ dependent, our system is of different nature than those studying the interplay of thermalization and Anderson localization \cite{cherroret15}.

\bigskip

{\it Kinetic equation.--}
It is important to recall that our experiments are carried out in the weakly nonlinear regime $L_{lin}\sim 1/\beta_0 \ll L_{nl} \sim 1/(\gamma N)$ \cite{baudin_arxiv}, and that 
linear propagation effects dominate disorder effects, $L_{lin} \ll L_d$ (or $\Delta \beta \ll \beta_0$).
According to this latter separation of spatial scales, turbulence in MMFs is described by a {\it discrete} wave turbulence approach \cite{PRL19,PRA19}, which means that only exact resonances contribute to the KE, while quasi-resonances can be neglected \cite{nazarenko11}.
Indeed, assuming that disorder effects dominate nonlinear effects $L_d \ll L_{nl}$, we have derived a {\it discrete} wave turbulence KE that describes the nonequilibrium evolution of the averaged modal components $n_p(z)=\left< |{\bm a}_p|^2(z)\right>$ during the propagation through the fiber \cite{PRL19,PRA19}:
\begin{eqnarray}
\nonumber
\partial_z n_p(z) &=&  \frac{ \gamma^2}{6\Delta \beta} \sum_{l,m,n} |S_{lmnp}|^2 \delta^K(\Delta\omega_{lmnp}) M_{lmnp}({\bm n}) \\
&&
+  \, \frac{4\gamma^2}{9 \Delta \beta}  \sum_l  
 |  s_{lp}({\bm n}) |^2 \delta^K(\Delta\omega_{lp})   (n_l-n_p), \quad \quad
\label{eq:kin}
\end{eqnarray}
with $s_{lp}({\bm n})=\sum_{m'} S_{lm'm'p} n_{m'}$, and $M_{lmnp}({\bm n})=  n_l n_m n_p+n_l n_m n_n -  n_n n_p n_m -n_n n_p n_l$ and $\Delta \omega_{lp}=\beta_l-\beta_p$.
The term $\delta^K(\Delta\omega_{lmnp})$ denotes the four-wave frequency resonance 
$\Delta \omega_{lmnp}  = \beta_l+\beta_m-\beta_n-\beta_p$, with 
$\delta^K(\Delta \omega_{lmnp})=1$ if $\Delta \omega_{lmnp}=0$, and zero otherwise.
Note the presence of $\Delta \beta$ in the denominator of the KE, so that disorder significantly affects the rate of thermalization \cite{PRA19}.

To derive the KE (\ref{eq:kin}) we made use of the conventional assumption of Gaussian statistics  to achieve a closure of the infinite hierarchy of the moments equations \cite{zakharov92}, a feature which is justified by the weakly nonlinear regime of our experiments. 
In the absence of the confining potential ($V(r)=0$), the wave turbulence KE can be derived under a weaker assumption than Gaussian statistics, namely the random phase and amplitude (RPA) approximation \cite{nazarenko11}. 
In the presence of the confining potential ($V(r) \neq 0$), we have shown that the Gaussian approximation gives the same result as the RPA for the coupling among non-degenerate modes, while differences appear for the degenerate modes, which only marginally affect the rate of thermalization. 

\begin{figure}
\includegraphics[width=1\columnwidth]{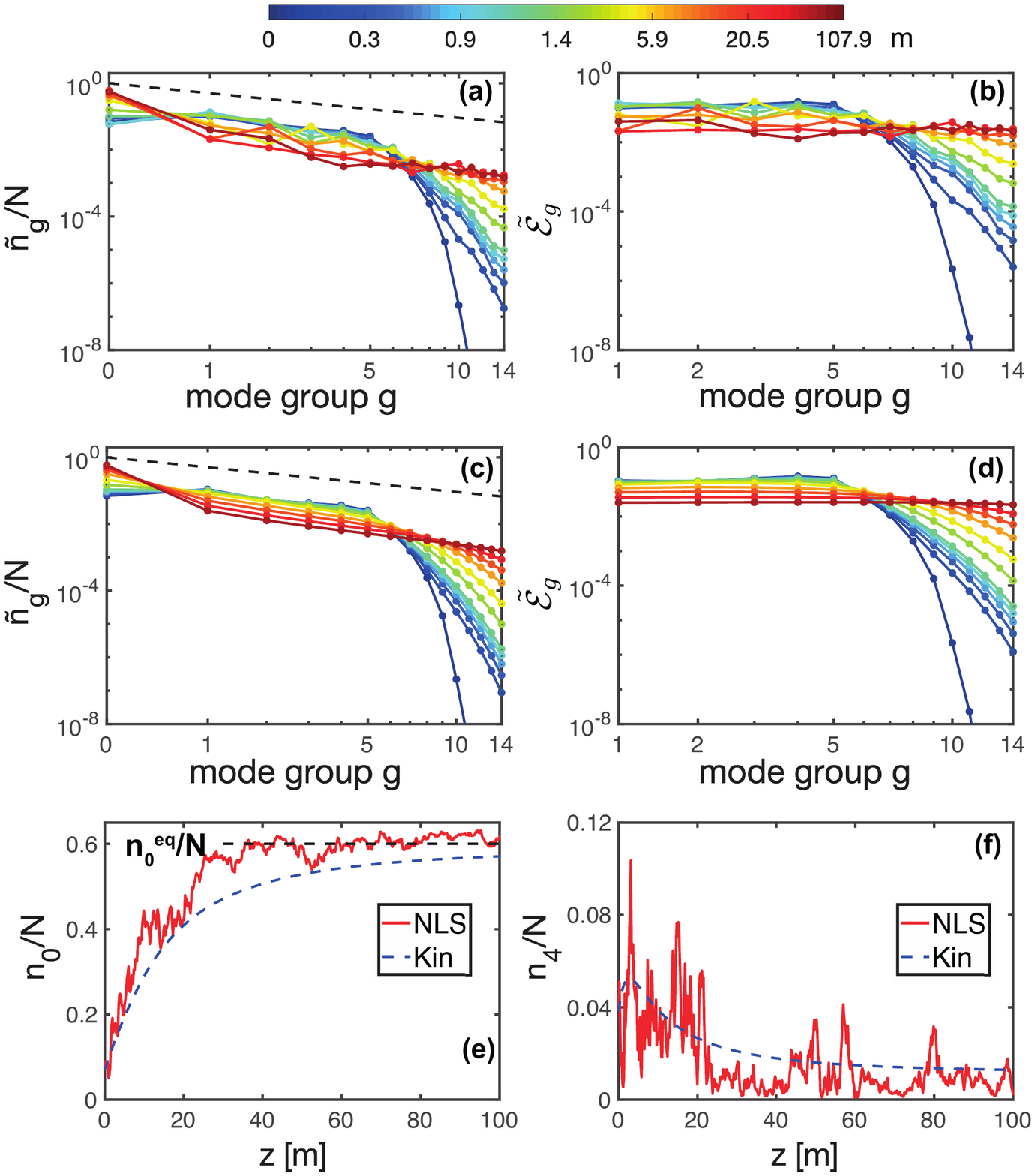}
\caption{
Numerical simulation of the modal NLSE (\ref{eq:nls}) (a)-(b), and KE (\ref{eq:kin}) (c)-(d):
Evolutions of the wave-action ${\tilde n}_g$ (a)-(c), and energy ${\tilde {\cal E}}_g$ (b)-(d), for  $g_{max}=$15 groups of non-degenerate modes.
The dashed black lines in (a) and (c) denote the RJ  power-law ${\tilde n}_g^{eq} \sim 1/g$.
The thermalization is featured by an energy flow toward the higher-order modes and a wave-action  flow toward the fundamental and higher-order modes.
Evolutions of $n_0(z)$ (e) and $n_4(z)$ (f) obtained from the NLSE (\ref{eq:nls}) simulation (red line) and the KE (\ref{eq:kin}) (dashed blue): The modal components thermalize to the theoretical equilibrium value predicted by the RJ theory (the dashed black line denotes $n_0^{eq}/N=0.6$).
Parameters: $N=$47.5kW, 
$\ell_c=0.019$m, $2 \pi /\sigma =0.26$m, there is no average over the realizations for the NLSE simulation.
}
\label{fig:1}
\end{figure}

\bigskip 
{\it Numerical simulations: Energy and wave-action flows.--}
The KE conserves the  wave-action $N=\sum_p n_p$ and the `energy' $E=\sum_p \beta_p n_p$ -- note that we call $E$ `energy' because it refers to the linear contribution to the Hamiltonian ($E$ is in units of W$\cdot$m$^{-1}$), while we call $N$ `wave-action' by following the wave turbulence terminology \cite{zakharov92} ($N$ is in units of W).
In a particle picture, $n_p$ and $N$ have the meaning of population of the $p$ mode and of total particle number.

The KE (\ref{eq:kin}) exhibits a $H-$theorem of entropy growth ($\partial_z {\cal S} \ge 0$) for the nonequilibrium entropy ${\cal S}(z)=\sum_p \log\big(n_p(z)\big)$, so that it describes an irreversible evolution to the RJ equilibrium distribution $n^{eq}_p=T/(\beta_p - \mu)$ that realizes the maximum of entropy.
Accordingly we have $N=T \sum_p (\beta_p-\mu)^{-1}$ and $E=T \sum_p \beta_p /(\beta_p-\mu)$ and we recall that there is a one to one relation between the equilibrium parameters $(\mu,T)$ and the initial conditions $(N,E)$ \cite{PRL05,PR14,christodoulides19b} -- note in particular that $T$ is not determined by a thermostat ($T$ is in units of W$\cdot$m$^{-1}$).

This irreversible process of thermalization to the RJ distribution is illustrated in Fig.~\ref{fig:1}(a)-(c), which reports  numerical simulations of the modal NLS Eq.(\ref{eq:nls}) and corresponding KE (\ref{eq:kin}) starting from the same initial condition.
During the propagation, $n_p$ essentially flows toward the fundamental mode (inverse cascade), while a small fraction of $n_p$ flows toward the higher-order modes.
For convenience we have reported in Fig.~1 the average wave-action ${\tilde n}_g$ within each group of degenerate modes, where $g=0,..,g_{max}-1$ indexes the mode group (in Fig.~1 $g_{max}=15$ for a total $M=g_{max}(g_{max}+1)/2=120$ modes).
The RJ power-law ${\tilde n}_g \sim 1/g$ is verified by the simulation of the KE and NLSE -- 
due to the large computation times, we are unable to perform an average over the realizations of NLSE simulations, which explains the noisy structure of ${\tilde n}_g$ in Fig.~1a-b.

These results are corroborated by the modal distribution of the energy, which exhibits a flow toward the higher-order modes (direct cascade).
In this example, we considered a relatively small value of the conserved energy $E$, which is below the critical value of the transition to condensation $E_{c} \simeq E_{min} \sqrt{M/2}$, where $E_{min}=N\beta_0$ denotes the minimum energy when all the `particles' $N$ populate the fundamental mode.
Note that $E_{c}$ only depends on the geometry of the waveguide potential, whose finite number of modes $M$ regularizes the ultraviolet catastrophe of classical waves.
In the condensed state, $\mu \to \beta_0^-$ \cite{baudin_arxiv}, so that the waves that started from an initial state with an excess energy in the low-energy modes, eventually tend to an equilibrium state displaying an energy equipartition among the modes ${\cal E}_p = (\beta_p-\beta_0) n_p \sim  T$ [or ${\tilde {\cal E}}_g = \beta_0 g {\tilde n}_g \sim  T$], as illustrated in Fig.~1(b)-(d).
Then RJ thermalization is characterized by a macroscopic population of the fundamental mode, as illustrated in Fig.~1(e), where the condensate fraction relaxes toward the theoretical equilibrium value $n_0^{eq}/N \simeq 0.6$. 
Note that the good agreement between NLSE and KE simulations in Fig.~1 is obtained {\it without using adjustable parameters}.

One may question whether the above energy and wave-action flows can be described theoretically by means of the Zakharov-Kolmogorov spectra of turbulence \cite{zakharov92}.
While these nonequilibrium stationary solutions are sustained by the addition of a  permanent forcing and damping at different scales in the system, they may be identified in the transient evolution of a purely conservative system, before reaching the RJ equilibrium \cite{nazarenko11,laurie12}. 
Note however that our KE (\ref{eq:kin}) differs from the conventional wave turbulence KE in two respects: 
(i) It involves the tensor $|S_{plmn}|^2$ instead of the Dirac $\delta-$function over the wave-vectors, because the potential $V(r)$ breaks the conservation of the momentum;
(ii) Our KE is discrete in frequencies. 
This latter property does not allow the application of the standard procedure based on the Zakharov conformal transformation to derive nonequilibrium stationary solutions featured by a non-vanishing flux of the conserved energy and wave-action.
This appears consistent with the numerical simulations, which  do not evidence the formation of a nonequilibrium power-law spectrum in the transient evolution that precedes the formation of the equilibrium RJ spectrum.

\bigskip

{\it Experimental results.--}
We performed experiments in a MMF with the experimental setup of Ref.\cite{baudin_arxiv}.
We used a 12m-long graded-index MMF that guides $M=120$ modes ($g_{max}=15$) with a core radius $R=26 \mu$m characterized by a parabolic shaped transverse refractive index.
The originality with respect to conventional experiments of spatial beam cleaning \cite{krupa16,wright16,krupa17} relies on the fact that the laser beam (Nd:YAG at $\lambda=1.06 \mu$m) is passed through a diffuser to generate a speckle beam before injection into the MMF. 
In the experiments we measure $N$ and $E$ from the near-field and far-field measurements of the intensity distributions, see Ref.\cite{baudin_arxiv}.
By moving the diffuser we obtain different realizations of the speckle beams, and then we can vary the (conserved) energy  $E$ while keeping constant the power ($N=7$kW).
Here, we focus the analysis into the near-field intensity distribution.
Note that, because of the parabolic shaped potential $V(r)$, the average near-field and far-field intensity representations are {\it equivalent} to each other \cite{baudin_arxiv}.

\begin{figure}
\centerline{\includegraphics[width=1\columnwidth]{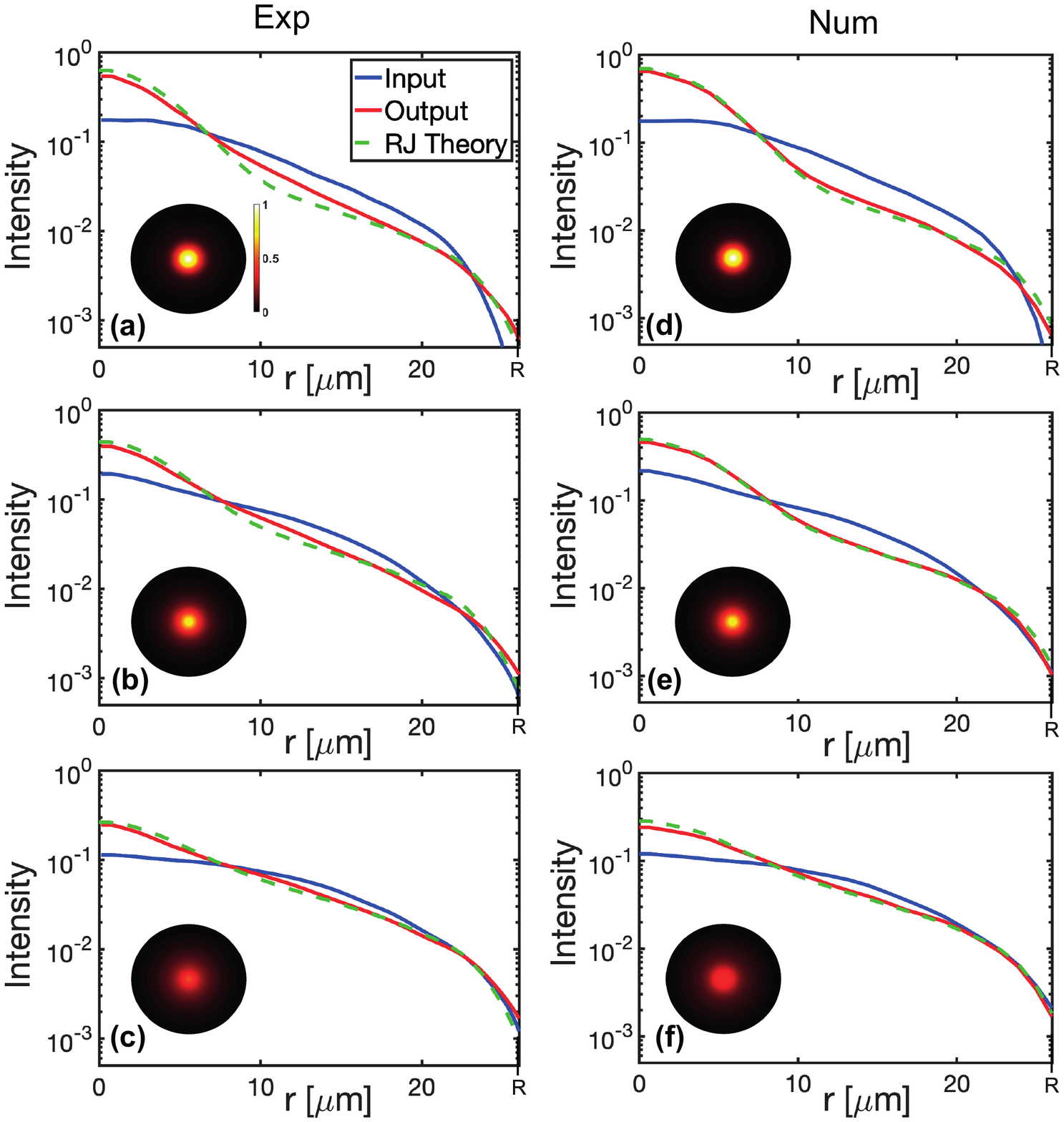}}
\caption{
(a)-(c) Experimental intensities distributions averaged over the realizations at the input (blue), and the output (red) of the MMF.
(d)-(f) Corresponding numerical simulations of the NLSE (\ref{eq:nls}), see the text for parameters.
The condensate fraction is $n_0^{eq}/N=0.6$ (1st line); $n_0^{eq}/N=0.4$ (2nd line); $n_0^{eq}/N=0.2$ (3rd line). 
The dashed green lines report the theoretical RJ intensity distribution $I^{eq}(r)$ from Eq.(\ref{eq:I_eq}) without using any adjustable parameter.
The intensities are plotted as a function of the angle-averaged distance $r=|\br|$.
The insets show the 2D output intensity distributions with the same color-bar (the circle denotes the MMF's core).}
\label{fig:2}
\end{figure}

We report in Fig.~2 (left column) the experimental results of the near-field intensity distributions  averaged over $\sim$50 realizations for three different values of the energies $E$, which correspond to an equilibrium condensate fraction of $n_0^{eq}/N=0.6, 0.4, 0.2$.
We report the `output' intensity distributions recorded at 12m (red lines), and the `input' intensities recorded after 20cm of propagation in the MMF (representing the `initial conditions' \cite{PRL19}, blue lines).
The output intensities are compared to the theoretical RJ intensity distributions $I^{eq}(r)$ (dashed green lines).
It is important to stress that the good agreement between the experiments and the theory  in Fig.~2 (left column) is {\it obtained without any adjustable parameter:} The experimentally measured values $(E,N)$ determine a unique pair $(\mu, T)$, 
which in turn determines $n_p^{eq}=T/(\beta_p-\mu)$ and thus the RJ equilibrium intensity distribution (dashed green lines in Fig.~2):
\begin{eqnarray}
I^{eq}(r)=\sum_p n_p^{eq} u_p^2(\br).
\label{eq:I_eq}
\end{eqnarray}
We do not have access to a measurement of the power $n_p$ within each individual mode $p$ in the experiments.
However, for large values of $p_x=p_y$, the asymptotic forms for the Hermite-Gauss functions show that the normalized mode $u_p(\br)$ is essentially supported in $r \le \sqrt{2g} r_o$ with $g=p_x+p_y$ \cite{szego}, i.e., there is  a correspondence between the radius $r$ and the mode number $g$.
The bi-directional wave-action flows toward the fundamental mode and the higher-order modes ($r \simeq R$) is clearly visible for a strong condensation, see  Fig.~2(a) for $n_0^{eq}/N=0.6$.
By increasing the energy $E$ (i.e. decreasing $n_0^{eq}/N$), the amount of incoherence (randomness) of the launched beam also increases and then populates the higher-order modes, so that only the inverse wave-action  flow toward the fundamental mode is  visible, see Fig.~2(c).
Note that, as recently demonstrated experimentally \cite{fabert20}, a self-cleaned optical beam exhibits a high degree of phase coherence.

The numerical simulations of the modal NLSE (\ref{eq:nls}) qualitatively reproduce the behavior observed experimentally.
This is illustrated in Fig.~2 (right column), where an average over the propagation has been considered from $12$m to $22$m so as to smooth the output intensity profiles (red lines). 
Although the parameters that characterize the disorder 
are not precisely known, we considered in Fig.~2 plausible experimental values $\ell_c=0.3$m and $2\pi/\sigma=2.14$m \cite{kaminow13}.
For these parameters disorder no longer dominates nonlinear effects ($L_d \sim L_{nl}$), and strictly speaking the KE (\ref{eq:kin}) is no longer valid \cite{PRA19}.
However, the scaling predicted by the KE, namely that thermalization is accelerated by decreasing the disorder (see the parameter $\Delta \beta$ in the denominator of (\ref{eq:kin})) is responsible for a fast process of condensation for the small disorder considered in Fig.~2.
This is apparent by comparing the simulations in Fig.~1 (propagated over $\sim 100$m) and Fig.~2 (over $L=12$m). 
In spite of the acceleration of thermalization, we had to increase the power up to 22kW  in the simulations to get a good agreement between NLSE simulations and the experimental results in Fig.~2. 
Then although the purely spatial model considered in Eq.(\ref{eq:nls}) captures many features of the experimental results, an improved quantitative agreement would require a spatio-temporal extension of the model so as to account for the pulsed laser regime considered in the experiments.

\bigskip

{\it Stability of the condensate.--}
The description of wave condensation in the absence of a trapping potential (i.e., in the homogeneous case $V(\br)=0$) is known to require a Bogoliubov approach, which shows that the condensate fraction $n_0^{eq}/N$ strongly depends on the nonlinearity $\gamma$ \cite{nazarenko05,PRL05}.
Here we show that the Bogoliubov approach is irrelevant to describe the weakly nonlinear regime of our experiment.

The structural disorder considered in the modal NLSE (\ref{eq:nls}) enforces the random phase dynamics among the modes.
As described by the KE (\ref{eq:kin}), the disorder then has a stabilizing effect on the process of condensation in the regime $L_d \ll L_{nl}$.
However, as discussed above through the simulations of Fig.~2, the disorder does not dominate nonlinear effects in the experiments.
In the following we show that the condensate is stable against the focusing nonlinearity even in the absence of disorder effects. 
Then we neglect the impact of polarization disorder and set ${\bf D}_p=0$, ${\bm a}_p \to a_p$ in Eq.(\ref{eq:nls}).
We assume that the fundamental mode is strongly occupied 
($|a_0| \gg |a_{m}|$, $m \neq 0$) and consider 
the weakly nonlinear regime $\eps = L_{lin}/L_{nl} = \gamma N/\beta_0 \ll 1$. 
The linearized equations read:
\begin{eqnarray}
\partial_z a_0 = -i \beta_0 a_0+ i \gamma  |a_0|^2 a_0 
\quad \quad \quad \quad \quad \quad \quad \quad \quad \nonumber \\
\quad \quad \quad \quad 
+ i \gamma  \sum_{p \neq 0}  s_{p0} (2 |a_0|^2 a_p +a_0^2 a_p^*) ,
\nonumber \\
\partial_z a_m = -i \beta_m a_m + i \gamma  s_{m0} |a_0|^2 a_0  
\quad \quad \quad \quad \quad \quad \quad \nonumber \\
\quad \quad \quad \quad 
+ i \gamma  \sum_{p \neq 0} s_{mp} (2 |a_0|^2 a_p +a_0^2 a_p^*),
\nonumber
\end{eqnarray}
where $s_{m n}=S_{mn00}$.
Writing $s_{m n}=w_{m_xn_x} w_{m_yn_y}$, we have 
 \begin{equation}
 w_{m_xn_x} = 
\frac{(-1)^{\frac{m_x-n_x}{2}}}{2^{m_x+n_x} \sqrt{m_x! n_x!}}
 \frac{(m_x+n_x)!}{(\frac{m_x+n_x}{2})!},
 \end{equation}
when $m_x$ and $n_x$ have the same parity, 
and $w_{m_xn_x}=0$ otherwise ({\it idem} for $w_{m_yn_y}$).
We look for a particular solution of the form $a_0=\sqrt{n_0} e^{- i \bar{\beta}_0  z }$,
where $\bar{\beta}_0$ will be defined later, and $a_m =d_m e^{- i   \bar{\beta}_0   z} $ 
with $\bar{\beta}_m = \beta_m - \bar{\beta}_0$ ($d_m$ real-valued).
The ansatz is solution if
\begin{eqnarray}
& - \bar{\beta}_0 n_0^{1/2}= - \beta_0 n_0^{1/2}+  \gamma    n_0^{3/2} + 3  \gamma n_0 \sum_{p \neq 0}  s_{p0}  d_p  \nonumber \\
& \bar{\beta}_m d_m = +  \gamma  s_{m0} n_0^{3/2}  
+ 3  \gamma  n_0 \sum_{p \neq 0} s_{mp}  d_p ,\quad m \neq 0  \nonumber 
\end{eqnarray}
Therefore the vector ${\bm d}$  is solution of the linear system $({\bf I} -{\bf K}){\bm d} = {\bm y}$, with 
the elements of ${\bm y}$ given by $y_m = \gamma n_0^{3/2} s_{m0}  /  \bar{\beta}_m $
and the matrix ${\bf K}=(K_{mp})$ given by 
$K_{mp} = 3 \gamma n_0 s_{mp}  / \bar{\beta}_m$ for $m\neq p$ and $0$ otherwise.
The matrix ${\bf I}-{\bf K}$ is invertible if 
$\sup_{m} \sum_{p} |K_{mp}| <1$, which is verified since $\eps \ll 1$.
Therefore there is a unique vector solution that is ${\bm d} = ({\bf I} -{\bf K})^{-1}  {\bm y}$.
By considering only the leading order corrections $O(\eps^2 \beta_0)$, we have $\displaystyle d_m = \gamma  n_0^{3/2} s_{m0} /( \beta_m -\beta_0)$, $\bar{\beta}_0=\beta_0 -\gamma  n_0$, $\bar{\beta}_m= \beta_m - \beta_0 + \gamma n_0$,   
and the nonlinear fundamental mode is of the form
$$
{\bar u}_0(\br, z) = \sqrt{n_0} e^{- i \bar{\beta}_0 z} \Big(u_0(\br) + \sum_{m \neq 0} 
 \frac{\gamma n_0  s_{m0} }{ {\beta}_m -\beta_0} u_m(\br)\Big)  .
$$
The field then consists of the superposition of the strong condensate  in the (slightly distorted) mode $\bar{u}_0$ and the incoherent mode fluctuations $a_{m \neq 0}$, that can be written in terms of $\tilde{a}_m(z) = a_m(z) e^{i \bar{\beta}_0 z}$:
\begin{eqnarray}
\label{eq:linearinc}
\partial_z \tilde{a}_m = -i \bar{\beta}_m \tilde{a}_m   
+ i \gamma n_0 \sum_{p \neq 0} s_{mp} (2  \tilde{a}_p +  \tilde{a}_p^*) .
\end{eqnarray}
The stability of this system is carried out by computing the matrix eigenvalues, which reveals that all eigenvalues are purely imaginary for $\eps \ll 1$, i.e., the condensate is stable.
Note that 
for $m_x,m_y \gg 1$, we have $s_{m0}\simeq 1/[\sqrt{\pi} (m_x m_y)^{1/4} 2^{(m_x+m_y)/2}]$ and $s_{mm} \simeq 4/[\pi (m_x m_y)^{1/2}]$, so that $s_{m0}$ exhibit a rapid decay to zero as compared to $s_{mm}$.
Then assuming $s_{mm} \gg s_{mp}$ ($p\neq m$), the eigenvalues are obtained in analytical form with the Bogoliubov dispersion relation
\begin{equation}
\label{def:km}
\bar{\beta}_{m}^{B}= \sqrt{( \bar{\beta}_m - 3 \gamma n_0 s_{mm} ) ( \bar{\beta}_m - \gamma n_0 s_{mm}  )} .
\end{equation}
Considering the weakly nonlinear regime of the experiment $\eps = \gamma N/\beta_0 < 10^{-3}$, 
$\bar{\beta}_{m}^{B}$ is real and $\bar{\beta}_{m}^{B} \simeq \bar{\beta}_m \simeq \beta_m-\beta_0$, i.e., the Bogoliubov dispersion relation of $a_m(z)$ in the presence of the condensate ($\sqrt{n_0} \gg |a_m|$) is well approximated by the linear expression $\beta_m^B \simeq \beta_m$.
In other words, the Bogoliubov nonlinear renormalization of the dispersion relation is negligible. 
This is corroborated by a scale-by-scale analysis of NLSE simulations, which reveals that even the strongly condensed mode $p=0$ evolves in the weakly nonlinear regime \cite{PRL19}.

\begin{figure}
\includegraphics[width=1\columnwidth]{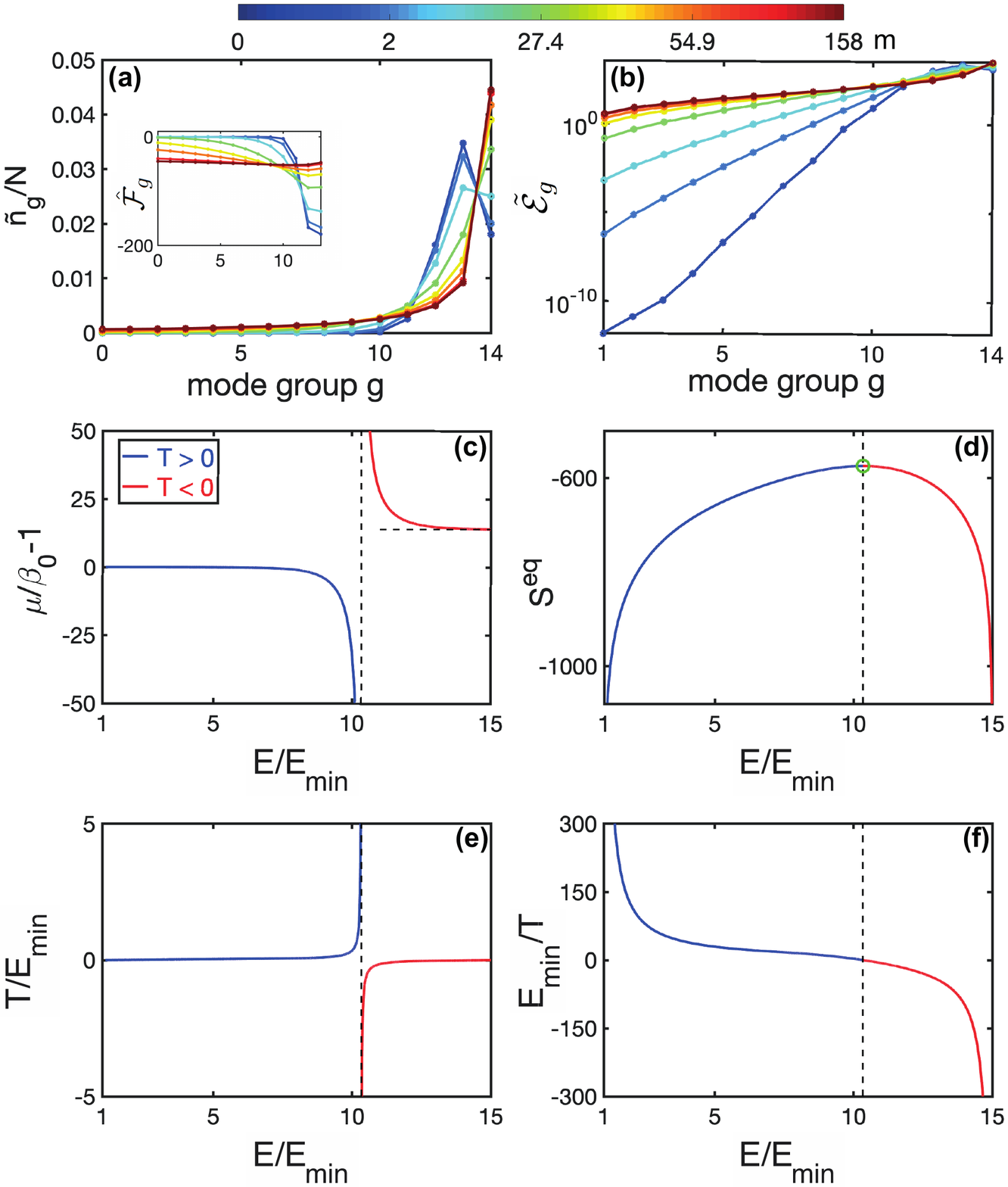}
\caption{
Simulation of the KE (\ref{eq:kin}) showing RJ thermalization toward a negative temperature equilibrium state:  ${\tilde n}_g$ essentially flows to the last group of degenerate modes (highest energy level $g=g_{max}-1$) (a), while the energy ${\tilde {\cal E}}_g$ flows to the low-order modes (b). 
The inset shows that ${\tilde {\cal F}}_g=\beta_0(g-g_{max}+1){\tilde n}_g$ relaxes toward an equipartition among the modes, ${\tilde {\cal F}}_g \simeq T < 0$ (red line), as predicted by the RJ equilibrium distribution
($N=$47.5kW, $\ell_c=0.3$m, $2 \pi /\sigma =2.1$m, $g_{max}=$15).
(c) $\mu/\beta_0 - 1$ vs $E/E_{min}$: note the asymptotic behaviors $\mu \to \beta_0^-$ for $E \to E_{min}$, and $\mu \to g_{max} \beta_0^+$ for $E \to E_{max}$. 
The horizontal dashed line denotes $\mu=g_{max} \beta_0$ and the vertical one $E=E_t$.
(d) $S^{eq}$ vs $E/E_{min}$ showing $1/T=(\partial S^{eq}/\partial E)_{M,N} < 0$ for $E > E_t$.
The green circle denotes $S^{eq}_{max}=-M \log M$ at $E=E_t$.  
(e) $T/E_{min}$ vs $E/E_{min}$: The divergences $T=\pm \infty$ for $E=E_t^{\mp}$ are removed  by plotting $E_{min}/T$ vs $E/E_{min}$ (f).
}
\end{figure}

\bigskip

{\it Perspectives on negative temperatures.--}
We have seen that 
light condensation in MMFs is driven by a flow of energy toward the higher order modes and a bi-directional redistribution of the wave-action.
This thermalization process exhibits properties similar to those identified numerically in the absence of a confining potential, see e.g. \cite{PRL18}.
However, the main difference is that condensation in a MMF is described by a weakly nonlinear and {\it discrete} wave turbulence approach where the structural disorder accelerates the process of thermalization over a relatively small number of modes ($M \simeq 120$), see the KE (\ref{eq:kin}).

An interesting consequence of the finite number of modes with an upper energy bound ($E_{max}=\beta_0 g_{max} N$) is that the system can exhibit negative temperature  equilibrium states, $T<0$ \cite{purcell,oja97,christodoulides19}.
The condition $n_p^{eq}=T/(\beta_p-\mu)>0$ then requires $\mu > {\rm max}(\beta_p) = g_{max} \beta_0$ and the equilibrium distribution is featured by an inverted modal population (${\tilde n}_{g+1}^{eq}>{\tilde n}_{g}^{eq}$) for an energy $E> E_t=N\left< \beta_p \right>=E_{min} (2g_{max}+1)/3$, where $\left< \beta_p \right>$ is the arithmetic mean of the eigenvalues and we recall that $E_{min}=N\beta_0$.
The denominator of the RJ equilibrium now vanishes for $\mu \to g_{max} \beta_0$.
Accordingly, ${\tilde n}_g$ essentially flows toward the {\it highest energy level}, i.e. highest mode group $g=14$, while the energy ${\tilde {\cal E}}_g=\beta_0 g {\tilde n}_g$ flows toward the {\it low-order modes}.
This process of thermalization toward a negative temperature equilibrium is demonstrated by the numerical simulation of the KE (\ref{eq:kin}) in Fig.~3.
For such a negative temperature equilibrium, the role of energy equipartition is played by the quantity ${\cal F}_p=(\beta_p-g_{max}\beta_0) n_p \simeq T <0$ [or ${\tilde {\cal F}}_g=\beta_0(g-g_{max}+1){\tilde n}_g \simeq T < 0$], which is equally distributed among the modes (inset of Fig.~3).
In spite of the fact that the highest energy level can be macroscopically populated ${\tilde n}_{g_{max}-1} \gg {\tilde n}_{g}$, there is {\it no phase coherence} amongst such a group of degenerate modes, which suggests an analogy with the notion of turbulent crystal \cite{newell93}.

We finally complete the study with the thermodynamic properties of the system.
We start from the {\it equilibrium} entropy 
${\tilde S}^{eq}=\sum_p \log(n_p^{eq})$ -- note that at equilibrium it coincides with the previous nonequilibrium entropy verifying the $H-$theorem. 
It proves convenient to shift the entropy by a constant ${S}^{eq}={\tilde S}^{eq}- M \log N$, so that by using  $T=N/\sum_p(\beta_p-\mu)^{-1}$, we can write
\begin{eqnarray}
{S}^{eq}(\mu)= - \sum_p\log ( \beta_p-\mu ) -M \log\Big( \sum_p \frac{1}{\beta_p-\mu} \Big) 
\label{eq:S_mu}\\
\frac{E(\mu)}{E_{min}} = \frac{ \sum_p \frac{\beta_p}{\beta_p-\mu}  }{ \sum_p \frac{\beta_0}{\beta_p-\mu}  }
\label{eq:E_mu}\\
\frac{T(\mu)}{E_{min}} = \frac{1}{\sum_p \frac{\beta_0}{\beta_p-\mu}}
\label{eq:T_mu}
\end{eqnarray}
The evolution of $\mu$ vs $E$ is reported in Fig.~3(c) from Eq.(\ref{eq:E_mu}).
It evidences that $\mu \to \beta_0^-$ for $E \to E_{min}$, and $\mu \to g_{max}\beta_0^+$ for $E \to E_{max}$: In both cases the denominator of the RJ distribution vanishes, which leads to the macroscopic population of the lowest mode ($g=0$) and the highest mode group ($g=15$), respectively.

The parametric plot with respect to $\mu$ of (\ref{eq:S_mu}) and (\ref{eq:E_mu}) gives ${S}^{eq}(E)$ in Fig.~3(d); while the corresponding parametric plot of (\ref{eq:E_mu}) and (\ref{eq:T_mu}) gives $T$ vs $E$ in Fig.~3(e).
Note the concavity of the entropy with respect of the energy as required by a self-consistent thermodynamic theory.
Negative temperatures equilibrium states arise for $E > E_t$, where the entropy {\it decreases} by increasing the energy, $T=(\partial E/\partial S^{eq})_{M,N} < 0$. 
Note that such negative temperature states ($E>E_t$) are actually ``hotter" than those at positive temperature ($E<E_t$), as the energy will spontaneously flow from negative to positive temperature when the systems are put in contact.

Remark in Fig.~3(e) that the equilibrium state corresponding to $T=0^+$ ($T=0^-$) refers to a population distribution concentrated in the lowest (highest) mode with $E=E_{min}$ ($E=E_{max}$).
Accordingly, the sates $T=0^+$ and $T=0^-$ are fundamentally different from each other, whereas there is almost no difference between the states $T=+\infty$ and $T=-\infty$ for $E \simeq E_{t}$.
This latter equilibrium state for $E = E_{t}$ corresponds to an equipartition of the wave-action among all the modes $n_p^{eq} =$const, and it refers to the most disordered state with $S_{max}^{eq}=-M \log M$ \cite{baudin_arxiv}, see the green circle in Fig.~3(c).
The apparent paradoxical divergence of $T=\pm \infty$ around this homogeneous state $n_p^{eq}=$const disappears if one considers the inverse of the temperature as the appropriate parameter (just as the Lagrange multiplier $1/T$ that arises naturally in statistical mechanics). 
In this case $1/T$ vs $E$ exhibits a continuous behaviour as shown in Fig.~3(f).

Work is in progress to study experimentally the unusual thermalization to negative temperature equilibrium states.
Given the large degeneracy of the condensate mode in this case, this raises interesting question about the possibility of having fragmented condensates~\cite{mueller06}.

\end{document}